\documentclass[twocolumn,showpacs,preprintnumbers,amsmath,amssymb,prb,nofootinbib]{revtex4}
\usepackage{amsfonts,amssymb,latexsym,amsmath,amscd}
\usepackage{times,mathptm,color}
\usepackage{anysize}
\usepackage{graphicx}
\usepackage[ps2pdf,colorlinks]{hyperref}

    {
    \smallskip
    \refstepcounter{theorem}
    \noindent
    {\bf Example \arabic{section}.\arabic{theorem}} \ \ }
    {\hspace*{\fill}{$\Diamond$}
    \smallskip}

    {
    \smallskip
    \refstepcounter{theorem}
    \noindent
    {\bf Remark \arabic{section}.\arabic{theorem}} \ \ }
    {\hspace*{\fill}{$\Diamond$}
    \smallskip}

    {
    \smallskip
    \refstepcounter{theorem}
    \noindent
    {\bf Definition \arabic{section}.\arabic{theorem}} \ \ }
    {\hspace*{\fill}{\ }
    \smallskip}

    {
    \noindent
    {\bf Proof{#1}:  }
    }
    {\hspace*{\fill}{$\Box$}\smallskip}

    {
    \noindent
    {\bf Sketch{#1}:  }
    }
    {\hspace*{\fill}{$\Box$}\smallskip}

    {
    \noindent
    {\bf Reference:  }
    }
    {\hspace*{\fill}{$\odot$}\smallskip}

\begin{document}
\bibliographystyle{apsrev}

\title{Theory of Rabi interaction between an infrared-active phonon and cavity-resonant modes}

\author{Hadley M. Lawler}
\altaffiliation[Current address: ]{Department of Physics, University of Washington, Seattle, Washington 98195, USA.}
\email{hmlawler@u.washington.edu}
\affiliation{Chemistry Division, Naval Research Laboratory, Washington, DC 20375, USA}
\author{Jason N. Byrd }
\affiliation{Department of Physics, University of Washington, Seattle, Washington 98195, USA}
\author{Gavin K. Brennen}
\affiliation{Institute for Quantum Optics and Quantum Information of the Austrian Academy of Sciences,
6020, Innsbruck, Austria \\}
\begin{abstract}
We present the theory of interaction between a polar vibration in a semiconductor and an electromagnetic 
mode of a surrounding cavity.  Tuning the cavity frequency near the transverse optical phonon frequency 
couples the phonon-induced polarization field within the dielectric and the electromagnetic field in 
the cavity.  Depending on engineering parameters, we predict that this cavity quantum electrodynamic 
interaction may reach the strong coupling regime.  The resonances of the system are in the terahertz spectral region, and while spectroscopic measurements are a possible route for the detection of 
such a system, we emphasize the possibility of measuring the Rabi oscillation directly in the time domain with femtosecond optical pulses.
\end{abstract}
\maketitle

\section{Introduction}

Weisbuch {\em et al.} initiated a new field of study with the demonstration of Rabi splitting between excitons and 
electromagnetic modes in nanoengineered semiconductor heterostructures\cite{1}, and similar phenomena have now been 
demonstrated in quantum-dot systems\cite{2}.   Fundamental and technologically promising developments are ongoing, 
including prospects for application to light-emitting diodes, probes of the quantum statistical limit\cite{3},
 and investigation of bosonic many-body effects\cite{4}.

Coincidentally, time-resolved studies of fast interactions in semiconductors have emerged with femtosecond laser 
technology\cite{5,6,7}. With new technology allowing time-domain, optical observations of fundamental excitations in 
crystals, coherent optical phonon oscillations have been read out in several systems, such as GaAs\cite{8} and 
bismuth\cite{9}.  Experiments have reported intriguing light-lattice interactions, such as phonon squeezing\cite{10},
pulse-controlled spatiotemporal lattice coherence\cite{11}, and exotic phonon dynamics\cite{12}.

We are proposing a measurement which may reveal a Rabi splitting between a coherent optical phonon in a polar crystal and 
a far-infrared cavity mode.  We mean "coherent" phonon to be taken in the contemporary sense, referring to a mode exhibiting
time-domain phase and amplitude information.  If such a Rabi splitting can be probed through a time-domain quantum beat, the phenomenon 
may be interpreted as a coherent oscillation of energy between a polar mechanical mode in the crystal and a vacuum cavity 
resonance.  As such, the system we suggest may demonstrate behavior which is interesting from the standpoint of 
narrow-band terahertz technology and quantum optics, and which presents favorable engineering parameters.  For instance, as 
will be discussed, a cavity-GaAs apparatus should exhibit a Rabi splitting of approximately 30 K.  This value roughly 
corresponds to the bulk longitudinal phonon frequency enhancement and the Kramers-Kronig-accompanied transverse phonon 
oscillator strength.  This robust Rabi coupling may be accessible with the relatively liberal requirement of 
several-micron-scale engineering.  

The Rabi interaction we are postulating is formally and physically similar to time-domain beating observed in two 
other coherent phonon systems.  At high pump fluences, a plasmon-mediated beating has been demonstrated\cite{13}, 
and more recently the phonon has been Rabi-coupled to superlattice-tuned excitonic states\cite{14}.    The intrinsic lifetime 
of the phonon mode and our estimate for the lifetime of the cavity mode are each greater than the fast and fluence-dependent 
dephasing time of the plasmon, and than the excitonic transition lifetime.  We shall discuss our view that the phonon and 
cavity-mode lifetimes will limit the coherent oscillation we propose.  For the photon mediated case we are addressing, this
suggests a more pronounced demonstration of strong-coupling behavior, and a greater duration of the oscillation.

With the above in mind, the qualitative features of our model system take shape.  The model possesses an in-plane 
translational invariance, so that the pump-driven optical phonon should maintain phase over in-plane lengths at least 
comparable to the dielectric thickness.  The effectively one-dimensional system remaining consists of a nodeless 
electromagnetic cavity mode superposed upon a uniform polarization, which follows a step function over the dielectric 
length.  The polarization is modeled as the combined effect of $N$ charged harmonic oscillators, one for each GaAs ionic 
pair.  Each oscillator has a natural frequency and a mass corresponding to the GaAs reduced mass, and a charge 
corresponding to the Born-effective charge of bulk GaAs.

Phononic oscillator strengths have been phenomenologically understood for many years\cite{15,16}, and it is no surprise 
that a coherently excited phonon may radiate. Terahertz radiation from pump-excited coherent phonons has indeed been 
observed\cite{17}.  The radiation spectra include an emission peak at the longitudinal-optical phonon frequency and a 
broadband response from the transient dynamics of photocarriers\cite{18}.  We are suggesting the possibility that a 
coherent phonon, in proximity to a cavity of well-defined resonance, can undergo a hybridization of its quantum state 
with the electromagnetic resonance, as is the case in atomic and excitonic systems.  In the weak-coupling limit, 
spontaneous photon emission and absorption are influenced, and in the strong-coupling limit emission and absorption 
can occur in coherent succession.  It may then be possible to control and enhance the amplitude and phase coherence 
of the excitations. In fact, it has now been experimentally confirmed that engineered geometries can enhance the 
terahertz radiation of coherent phonons by two orders of magnitude\cite{19}, and that the polaritonic spectrum can 
be tuned by varying electromagnetic boundary conditions\cite{20}.

If the phenomena we describe can be observed, and optimally controlled, they may provide a promising specimen 
for the investigation of macroscopic quantum phenomena.  They may also prove relevant to important new terahertz 
technologies.  For instance, the system we describe may allow more detailed study the phonon signal limiting 
electro-optic sampling, and is related to the control of terahertz polaritonic waveforms and phonon-polariton photonics
\cite{21,22}.
The amplitude decay may also help to illuminate the nature of the coherent state, and probe interactions among 
lattice, electronic and electromagnetic degrees of freedom.

We will refer to the coupled phonon-vacuum resonance excitation as a phononic cavity polariton.  Traditionally, 
the polariton was considered as a bulk excitation\cite{9}, but there is a conceptual resemblance between the 
canonical polariton dispersion and a cavity-resonant system.  If one imagines a level (avoided) crossing diagram 
representing the variation of the mechanical and electromagnetic energies with inverse cavity dimension, the 
result is not unlike the standard polariton dispersion, with the correspondence of wave vector to inverse cavity 
dimension.  In this picture, the mechanical resonance is constant at the transverse-optical phonon frequency, 
while the electromagnetic resonance energy increases linearly with the inverse dimension.  Thus Fig. 2 
bears resemblance to the polariton dispersion of bulk GaAs.  

Accordingly, Savona {\em et al.}, in a study of excitonic cavity polaritons, generalized Hopfield's bulk 
polariton theory to allow for a vacuum cavity, and for the extension of the electromagnetic field beyond the 
boundaries of a confined condensed-matter system\cite{23}.  In this work, we develop the analogous theory for 
the phononic cavity polariton.

\section{Derivation of Interaction Hamiltonian}

\begin{figure}
\begin{center}
\includegraphics[width=\columnwidth]{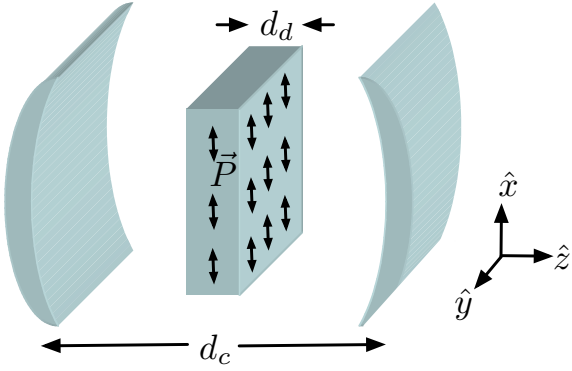}
\end{center}
\caption{\label{fig:1}Schematic of a dielectric crystal with polarization current $\vec{P}=P\hat{y}$ inside a cavity.  The length scales satisfy $d_c\approx c\pi/2\omega_{TO}$ and $d_d<d_c\ll l_x,l_y$ where 
 $l_{x,y}$ are the transverse dimensions of the dielectric.}
\end{figure}

The formalism to follow will emulate Savona's light-matter interaction theory for exciton polaritons, which 
motivated it.  Our system is comprised of a rectangular, semiconducting crystal, of thickness $d_d$, inside a 
cavity bound by far-infrared reflective surfaces.  The half-wave resonance length, including the dielectric 
and the cavity, is $d_c$.  In realizing such a system, a convenience is presented by the high reflectivity 
of simple metals in the far infrared.  This property makes the use of Bragg reflectors and other advanced 
micro-fabrication techniques, which are required at optical frequencies, unnecessary in the far-infrared. The 
semiconductor providing lattice motion must lack an inversion symmetry, so that it possesses an optically 
active phonon.  The simplest, and best studied such material is GaAs, and so we will adopt it in the theory 
to follow.  The geometry and relevant lengths are illustrated in Fig.~\ref{fig:1}.


The far-infrared frequencies are much smaller than the fundamental electronic gap of GaAs, and so, neglecting 
any photocarriers, our Hamiltonian need only consider the bare piece $H_0$ which includes the phonon field in the lattice and the electro-dynamic field, and their 
interaction, i.e.
\begin{equation}
H=H_0+ H_{\rm I}.
\end{equation}
Generally, the lattice term is a set of harmonic oscillators, each with a wavevector within the Brillouin 
zone, and there are three branch index values per atom in the primitive unit cell.  Similarly, the photon field 
can be expanded into components of given wavevector and polarization, such that the reflective boundary 
condition is maintained.  In the crudest approximation, we limit our consideration of the electro-dynamic 
degrees of freedom to the lowest-energy mode satisfying the reflective boundary condition, the half-wave 
resonance.  The general multimode case is treated in the Appendix.  Furthermore, we limit the lattice-light interaction to lattice modes of vanishing crystal momentum, 
relative to Brillouin zone dimensions.  This is justified from energy and momentum conservation in the emission 
and absorption of photons by lattice vibrations\cite{24}. Finally, both the phonon and photon polarizations 
are taken to be in the $\hat{x}$-direction, which is within the plane of the interface and reflective surfaces.

With these approximations in mind, the photon and phonon contributions can be written explicitly in terms 
of the field creation and annihilation operators,
\begin{equation}
H=\hbar \omega_{TO}(a^\dagger a+1/2)+\frac{\pi \hbar c}{\sqrt{\epsilon}d_c}(b^\dagger b+1/2)+H_{\rm I}.
\end{equation}
The first term on the right hand side is the lattice term, and indicates the transverse optical phonon frequency.   The second term is the photon term, and indicates the speed of light and the effective dielectric constant $\epsilon$, which 
characterizes the composite system of cavity and dielectric.  A reasonable approximation to $\epsilon$, which will be assumed here, is a volume average of the bulk semiconductor dielectric 
constant and the permittivity of free space; i.e. 
\begin{equation}
\epsilon=1+(\epsilon_0-1)d_d/d_c.
\end{equation}
A more accurate value may be obtained through meeting Maxwell's 
boundary conditions at the interfaces and reflective surfaces, but such a correction only shifts the cavity 
resonance, and we are tuning near the transverse optical phonon frequency,  $\omega_{TO}$, in any event. In 
the second term on the right, the creation and destruction operators correspond to photons with polarization, 
again, along the $\hat{x}$-axis.

The derivation of our interaction Hamiltonian closely resembles Savona $\emph{et al's}$ excitonic theory.  The general form 
of the interaction, with electromagnetic vector potential $A$ polarized along $\hat{x}$, is:
\begin{equation}
H_{\rm I}=\frac{-Z}{c}\sum_r A(r) \frac{1}{i\hbar}[x(r),H_{\rm p}] + \frac{Z^2}{2\mu c^2} \sum_r A^2(r).
\end{equation}
This sum is taken over the lattice site positions, where each Ga and As atomic pair are interpreted as 
harmonically oscillating dipoles, characterized by Born-effective charge, $Z$, reduced mass, $\mu$, and frequency
$\omega_{TO}$.  The x-projection of the normalized, local, harmonic oscillator displacement field can be 
expressed through the phonon operators above:
\begin{equation}
x(r)=\frac{1}{\sqrt{N}}\sqrt{\frac{\hbar}{2\mu\omega_{TO}}}(a+a^\dagger),
\end{equation}
where there are $N$ atomic pairs within the dielectric. The commutator in the interaction Hamiltonian, then, is:
\begin{equation}
\frac{1}{i\hbar}[x(r),H_{\rm p}]=-i\sqrt{\frac{\hbar\omega_{TO}}{2N\mu}}(a-a^\dagger).\\
\end{equation}

Now, the normalized vector potential is approximated as a planar, standing halfwave, polarized along 
$\hat{x}$.  The phase gradient is along the interface normal, $z$, and is approximated by its 
length-averaged value, $q=\pi/d_c$. If the dielectric and the cavity share a cross-sectional area $S$, then, 
the $\hat{x}$-polarized component of the vector potential is:
\begin{equation}
A(r_z)=\left(\frac{2\pi \hbar c}{Sd_cq\sqrt{\epsilon}}\right)^{1/2} \cos[qr_z](b+b^\dagger).
\end{equation}

The first term in the interaction Hamiltonian may be re-written as:
\begin{equation}
\begin{array}{lll}
\frac{-Z}{c}\sum_r A(r)\frac{1}{i\hbar}[x(r),H_{\rm p}]&=& 
\frac{iZ}{c} \sum_r \left(\frac{2\pi \hbar c}{S\pi \sqrt{\epsilon}}\right)^{1/2}
\sqrt{\frac{\hbar\omega_{TO}}{2N\mu}} \\
&&
\cos[qr_z](a-a^\dagger)(b+b^\dagger). 
\end{array}
\end{equation}
The sum over atomic pairs, each occupying a volume $\Omega$, may be approximated with an integral over the z direction:
\begin{equation}
\sum_r \cos[qr_z]=\frac{S}{\Omega}\int_{-d_d/2}^{d_d/2}dr_z \cos[qr_z].
\end{equation}
The final form of the first term in the interaction Hamiltonian follows,
\begin{equation}
\begin{array}{lll}
\frac{-Z}{c} \sum_r A(r)\frac{1}{i\hbar}[x(r),H_{\rm p}]&=&\frac{ih \omega_p}{\pi^{3/2}} \sqrt{\frac{\omega_{TO}}{c\sqrt{\epsilon}}}\frac{d_c}{\sqrt{d_d}}\\
&&\sin[\frac{\pi d_d}{2d_c}]
(a-a^\dagger)(b+b^\dagger)
\end{array}
\end{equation}
Where $N\Omega=S d_d$, and the conventional definition of the 
nuclear plasma frequency is substituted, $\omega_p=Z(4\pi/\mu\Omega)^{1/2}$\cite{16}.
The second term in the interaction Hamiltonian is:
\begin{equation}
\begin{array}{lll}
\sum_r A^2(r)&=&\sum_r \frac{Z^2}{2\mu c^2}\frac{2\pi \hbar c}{Sqd_c\sqrt{\epsilon}}\cos^2[qr_z](b+b^\dagger)^2\\
&=&\frac{\omega_p^2}{8\pi}\left(\frac{\hbar}{c\sqrt{\epsilon}}\right)
\left(d_d+\frac{d_c}{\pi}\sin\Big[\frac{\pi d_d}{d_c}\Big]\right)(b+b^\dagger)^2.
\end{array}
\end{equation}

With these explicit expressions for the interaction term in the Hamiltonian, the whole Hamiltonian for the system is,
\begin{equation}
\begin{array}{lll}
H&=&\hbar\omega_{TO}(a^\dagger a+1/2)+\frac{\hbar c \pi}{\sqrt{\epsilon}d_c}(b^\dagger b+1/2)\\
&&+i\hbar \omega_{AP}(b+b^{\dagger})(a-a^{\dagger})\\
&&+\hbar\omega_{AA}(b+b^\dagger)^2.
\end{array}
\label{effHam}
\end{equation}
The interaction coefficients are taken from above,
\begin{equation}
\begin{array}{lll}
\omega_{AP}&  =& \frac{\omega_p}{\pi^{3/2}}\sqrt{\frac{\omega_{TO}}{c\sqrt{\epsilon}}}\frac{d_c}{\sqrt{d_d}}
\sin[\frac{\pi d_d}{2d_c}],\\
\omega_{AA}&  =& \frac{\omega_p^2}{8\pi}\left(\frac{1}{c\sqrt{\epsilon}}\right)
\left(d_d+\frac{d_c}{\pi}\sin[\frac{\pi d_d}{d_c}]\right).
\end{array}
\label{freqs}
\end{equation}
Physically, this interaction corresponds to a uniform polarization current coupling to the positive and negative wavevectors of the lowest frequency standing wave that matches the boundary conditions.

\section{Near Resonant Behavior}

Simplistic physical considerations lead to the conclusion that any eigenstate of our system requires a 
mixture of lattice and electromagnetic character.  A  phonon will radiate, and conversely a photon will 
be absorbed by the crystal, so that neither one of these states alone is stationary.  To find an 
eigenstate of the Hamiltonian, which we refer to as the cavity polariton, we define its operator subject 
to two conditions.  First, we require the polariton to be some superposition of phonon and photon fields; 
and second, we require that the commutator of the polariton operator with the Hamiltonian take the standard 
harmonic form.

The former is satisfied by writing the polariton field operator as a linear combination of creation and 
annihilation operators of the phonon and photon fields, 
\begin{equation}
B=Wa+Xb+Y a^\dagger + Z b^\dagger.
\end{equation}
Each of the coefficients, $W$, $X$, $Y$, and $Z$, implicitly depends on the cavity dimension, $d_c$, through the
couplings, $\hbar\omega_{AP}$ and $\hbar\omega_{AA}$. The second requirement is expressed as,
\begin{equation}
[B,H]=EB,
\end{equation}
where on the right-hand side the energy eigenvalue is indicated.

\begin{figure}
\begin{center}
\includegraphics[width=\columnwidth]{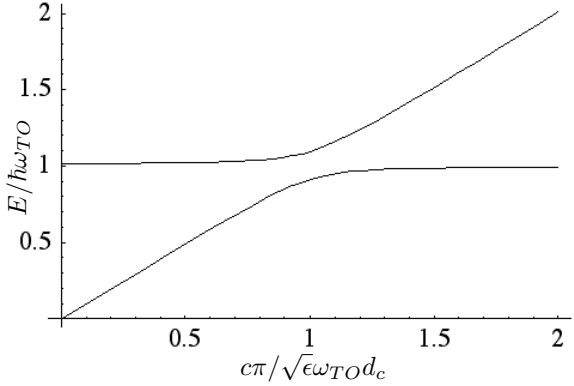}
\end{center}
\caption{\label{fig:2}Variation of the positive energy eigenstates of the polariton Hamiltonian as a function of the inverse cavity length $d_c$.  The dielectric is taken to be a GaAs crystal of width $d_d=d_c/50$.  The resonance occurs near $ d_c\sim\pi c/\sqrt{\epsilon}\omega_{TO}=16.45\mu $m.}
\end{figure}

Making the substitutions, $D=\hbar\omega_{AA}$, and $C=\hbar\omega_{AP}$, 
the above two equations may be combined in matrix form:
\begin{widetext}
\[
\left[
\begin{array}{cccc}
\frac{\hbar \pi c}{\sqrt{\epsilon}d_c}+2D-E & -iC & -2D & -iC \\
iC & \hbar\omega_{TO}-E & -iC & 0 \\
2D & -iC & -\frac{\hbar \pi c}{\sqrt{\epsilon}d_c} -2D-E & -iC \\
-iC & 0 & iC& -\hbar\omega_{TO}-E \\
\end{array}
\right]
\left(
\begin{array}{c}
W\\X\\Y\\Z
\end{array}
\right)
=0
\]
\end{widetext}
The eigenvalues are obtained by solving for the condition that the above $4\times4$ matrix have zero determinant.  The exact expressions for positive branch of energies that exhibit an anticrossing is:
\begin{equation}
\begin{array}{lll}
E_{\pm}&=&\hbar \Big[(\omega^2+4\omega_{AA}\omega+\omega_{TO}^2\pm (\omega^4+
8\omega_{AA}\omega^3\\
&&+16\omega_{AA}^2\omega^2-2\omega_{TO}^2\omega^2-8\omega_{AA}\omega_{TO}^2\omega\\
&&+16\omega_{AP}^2\omega_{TO}\omega+\omega_{TO}^4)^{1/2})/2\Big]^{1/2}
\end{array}
\end{equation}
where $\omega=c\pi/\sqrt{\epsilon}d_c$.
These energies, as a function of $d_c$, are plotted in Fig. \ref{fig:2}.  When the cavity is tuned far from $\omega_{TO}$, 
the two branches have negligible curvature, with the slope of the photon-like branch proportional to $c$, and the 
slope of phonon-like branch approximately zero.  In this limit, the cavity mode and lattice vibration are 
decoupled.  When the cavity mode is tuned near $\omega_{TO}$, however, there is a significant coupling, 
and there is evident curvature in the two branches. 

The Rabi frequency at maximum coupling can be seen in Fig. \ref{fig:2} as the minimum value of the energy difference between the two branches.  The figure also indicates the cavity detuning range over which 
significant coupling may be expected.  The parameters for GaAs are $\omega_{TO}=36$ microns, $\omega_{p}=51$ microns, and $\epsilon_0=10.89$.
The crystal width to cavity width ratio is taken to be fixed at $d_d=d_c/50$.

\section{Dephasing and Prospects for Measuring Rabi Oscillation}

When a cavity-mode is weakly coupled to an excitation of non-zero oscillator strength, spontaneous emission 
and absorption rates can be influenced by tuning the cavity through the resonance.  When the coupling is 
increased, and the inverse energy difference of the avoided crossing surpasses a characteristic dissipation 
rate, coherent interaction may be observed.  For instance, a Rabi oscillation can take place between modes 
of primarily lattice character, and of primarily electromagnetic character.  The fundamental distinction 
between the strong- and weak-coupling conditions is governed by the relative magnitudes of the near-resonant 
oscillator strength of the excitation, and the mode and excitation lifetimes. 

Near cavity resonance, the strong coupling condition is $\omega_{AP}>|\gamma_c-\gamma_{TO}|/4$, where $\gamma_c,\gamma_{TO}$ are the full-width at half-maximum linewidths for the cavity mode and phonon \cite{25}.  For $d_d\ll d_c$ this implies
\begin{equation}
w_p \sqrt{d_d/d_c} > |\gamma_c - \gamma_{TO}|/2.
\end{equation}

The bulk GaAs phonon linewidth is well-studied\cite{24}, and is in the thousands of microns.  The cavity 
lifetime is a less fundamental quantity, and depends on engineering of the cavity.  State-of-the-art 
etalon techniques can increase the cavity lifetime at the expense of the tuning range.  With a two-micron 
tuning range, a cavity lifetime which is again in the thousands of microns is possible with commercially 
available technology.  We see than that the above inequality is satisfied to an order of magnitude.  This strong coupling is a consequence of the two-dimensional emitter at the interface.

We conclude then that strong coupling is likely.  The Rabi period is also greater than the 
temporal resolution of ultrafast techniques using femtosecond optical pulses.  Presuming that the 
dependence of the reflectivity on lattice strain is distinguishable from the effect of the electromagnetic 
field, the Rabi oscillation should be observable by pumping the optical phonon, and performing time-domain 
reflectivity measurements.

\section{summary and conclusions}

We express  the quantum interaction between an optically active phonon  and a surrounding cavity's resonant 
electromagnetic mode.  When the two are tuned, we demonstrate a Rabi splitting, which is evident in the calculated 
energy levels for varying cavity dimension.  Furthermore, we review various experimental studies and analyze the 
strong-coupling criteria to support our view that a time-domain Rabi oscillation may be observed in the system we describe.

\acknowledgements
We are grateful to Sanjiv Shresta, whose comments and insights led to our consideration of the work.  H.L. received support from the National Research Council through a NRC-NRL Resident Research Associateship and from the Office of Naval Research both directly and through the Naval Research Laboratory.  JNB thanks the NSF REU program.  GKB received support from the Austrian Science Foundation and the Institute for Quantum Information.

\appendix
\section{}
In the following, we present a full multimode calculation of the the polariton Hamiltonian and the  dispersion relation for a dielectric crystal embedded in a cavity.  The bare Hamiltonian for our system is (up to an overall constant):
\begin{equation}
H_0=\sum_{\vec{q},\sigma}\hbar \omega_{TO}a^{\dagger}_{\vec{q},\sigma}a_{\vec{q},\sigma}+\hbar \frac{c}{\sqrt{\epsilon}} |\vec{q}|b^{\dagger}_{\vec{q},\sigma}b_{\vec{q},\sigma}.
\end{equation} 
Here $a_{\vec{q},\sigma},a^{\dagger}_{\vec{q},\sigma}$ are the annihilation and creation operators for the phonon field at the frequency $\omega_{TO}$.  The phonon wave vector $\vec{q}=(\vec{q}_{\perp},q_z)$ is taken to be small (so that wave lengths are macroscopic), and the phonon polarization vector is $\hat{e}_{\sigma}$. Similarly, $b_{\vec{q},\sigma},b^{\dagger}_{\vec{q},\sigma}$ are the annihilation and creation operators for the photon field.  The dielectric constant is the volume averaged value $\epsilon=1+(\epsilon_0-1) d_d/d_c$.  In the radiation gauge, the Hamiltonian describing the interaction between the fields is:
\begin{equation}
\begin{array}{lll}
H_I&=&-\frac{1}{c}\displaystyle{\int_{crystal}}d^3 x \vec{A}\cdot \dot{\vec{P}}+  \frac{Z^2}{2\Omega\mu c^2} \displaystyle{\int_{crystal}} d^3x\vec{A}\cdot\vec{A}\\
&=&H_I^{(1)}+H_I^{(2)}.\;
\end{array}
\end{equation}
where $\Omega$ is the volume per lattice cell of atomic pairs, $\mu$ is the reduced mass of the pair, $Z$ is the Born effective charge, and $c$ is the vacuum speed of light.  The macroscopic polarization current is
\begin{equation}
\begin{array}{lll}
\vec{P}&=&\sqrt{\frac{\hbar Z^2}{2\Omega\mu\omega_{TO}Sd_d}}\displaystyle{\sum_{q_y,q_z}}(a_{q_y,q_z}e^{i(q_y y+q_zz)}\\
&+&a^{\dagger}_{q_y,q_z}e^{-i(q_y y+q_zz)})\hat{x}.
\end{array}
\end{equation}
The vector potential must satisfy appropriate boundary conditions along the cavity axis $\hat{z}$.  We assume perfectly reflecting mirrors at the cavity boundaries meaning the momentum components of the field along $\hat{z}$ satisfy $q_z=n\pi/d_c$ for $n\in \mathbb{Z}$.  The suitably normalized vector potential is then
\vbox{
\begin{equation}
\begin{array}{lll}
\vec{A}&=&\sqrt{\frac{2\pi \hbar c}{\sqrt{\epsilon}S d_c}}\displaystyle{\sum_{q_x,q_y}\sum_{n\in\mathbb{Z}}}\Big(\frac{\sin(2n\pi z/d_c)}{(|\vec{q_{\perp}}|^2+(2n\pi/d_c)^2)^{1/4}}\\
&+&\frac{\cos((2n+1)\pi z/d_c)}{(|\vec{q_{\perp}}|^2+((2n+1)\pi/d_c)^2)^{1/4}}\Big)\\
&\times&\sum_{\sigma}(b_{\vec{q},\sigma}e^{i\vec{q}_{\perp}\cdot \vec{x}}+b^{\dagger}_{\vec{q},\sigma}e^{-i\vec{q}_{\perp}\cdot \vec{x}})\hat{e}_{\sigma}.
\end{array}
\end{equation}}
Hereafter we suppress the polarization index $\sigma$ as it is understood that only $\hat{x}$ polarized fields contribute to the interaction.  Using the relation $\dot{\vec{P}}=-\frac{i}{\hbar}[\vec{P},H_0]$, we obtain the full expression for the interaction terms:
\begin{widetext}
\begin{equation}
\begin{array}{lll}
H_I^{(1)}&=&i\hbar\omega_p\sqrt{\frac{ \omega_{TO}}{4cS^2d_cd_d\sqrt{\epsilon}}}\displaystyle{\int_{crystal}d^3x \sum_{q_y,q'_y,q'_z}\sum_{n\in\mathbb{Z}}}\Big(\frac{\cos((2n+1)\pi z/d_c)}{(q^2_y+((2n+1)\pi/d_c)^2)^{1/4}}(b_{q_y,\frac{(2n+1)\pi}{d_c}}a_{q'_y,q'_z}e^{i(q_y+q'_y)y}e^{iq'_z z}\\
&-&b_{q_y,\frac{(2n+1)\pi}{d_c}}a^{\dagger}_{q'_y,q'_z}e^{i(q_y-q'_y)y}e^{-iq'_z z}+b^{\dagger}_{q_y,\frac{(2n+1)\pi}{d_c}}a_{q'_y,q'_z}e^{-i(q_y-q'_y)y}e^{iq'_z z}-b^{\dagger}_{q_y,\frac{(2n+1)\pi}{d_c}}a^{\dagger}_{q'_y,q'_z}e^{-i(q_y+q'_y)y}e^{-iq'_z z})\\
&+&\frac{\sin(2n\pi z/d_c)}{(q^2_y+(2n\pi/d_c)^2)^{1/4}}(b_{q_y,\frac{2n\pi}{d_c}}a_{q'_y,q'_z}e^{i(q_y+q'_y)y}e^{iq'_z z}-b_{q_y,\frac{2n\pi}{d_c}}a^{\dagger}_{q'_y,q'_z}e^{i(q_y-q'_y)y}e^{-iq'_z z}+b^{\dagger}_{q_y,\frac{2n\pi}{d_c}}a_{q'_y,q'_z}e^{-i(q_y-q'_y)y}e^{iq'_z z}\\
&-&b^{\dagger}_{q_y,\frac{2n\pi}{d_c}}a^{\dagger}_{q'_y,q'_z}e^{-i(q_y+q'_y)x}e^{-iq'_z z})\Big)\\
&=&\displaystyle{\sum_{q_y,q'_z}\sum_{n\in\mathbb{Z}}}\Big(-C^o_{(q_y,\frac{2n\pi}{d_c}),q'_z}(b_{q_y,\frac{2n\pi}{d_c}}a_{-q_y,q'_z}
+b_{q_y,\frac{2n\pi}{d_c}}a^{\dagger}_{q_y,q'_z}+b^{\dagger}_{q_y,\frac{2n\pi}{d_c}}a_{q_y,q'_z}+b^{\dagger}_{q_y,\frac{2n\pi}{d_c}}a^{\dagger}_{-q_y,q'_z})\\
&+&iC^e_{(q_y,\frac{(2n+1)\pi}{d_c}),q'_z}(b_{q_y,\frac{(2n+1)\pi}{d_c}}a_{-q_y,q'_z}
-b_{q_y,\frac{(2n+1)\pi}{d_c}}a^{\dagger}_{q_y,q'_z}+b^{\dagger}_{q_y,\frac{(2n+1)\pi}{d_c}}a_{q_y,q'_z}-b^{\dagger}_{q_y,\frac{(2n+1)\pi}{d_c}}a^{\dagger}_{-q_y,q'_z})\Big)\\
H_I^{(2)}&=&\frac{ \hbar\omega_p^2 }{4c\sqrt{\epsilon}d_cS}\displaystyle{\int_{crystal}d^3x\sum_{q_y,q'_y}\sum_{m,n\in\mathbb{Z}}} \Big(\frac{\cos((2m+1)\pi z/d_c)\cos((2n+1)\pi z/d_c)}{(q^2_y+((2m+1)\pi/d_c)^2)^{1/4}(q^2_y+((2n+1)\pi/d_c)^2)^{1/4}}\\
&\times&(b_{q_y,\frac{(2m+1)\pi}{d_c}}e^{iq_y y}+b^{\dagger}_{q_y,\frac{(2m+1)\pi}{d_c}}e^{-iq_y y})
(b_{q_y,\frac{(2n+1)\pi}{d_c}}e^{iq'_y y}+b^{\dagger}_{q_y,\frac{(2n+1)\pi}{d_c}}e^{-iq'_y y})
+\frac{\sin(2m\pi z/d_c)\sin(2n\pi z/d_c)}{(q^2_y+(2m\pi/d_c)^2)^{1/4}(q^2_y+(2n\pi/d_c)^2)^{1/4}}\\
&\times&(b_{q_y,\frac{2m\pi}{d_c}}e^{iq_y y}+b^{\dagger}_{q_y,\frac{2m\pi}{d_c}}e^{-iq_y y})
(b_{q_y,\frac{2n\pi}{d_c}}e^{iq'_y y}+b^{\dagger}_{q_y,\frac{2n\pi}{d_c}}e^{-iq'_y y})\\
&=&\displaystyle{\sum_{q_y}\sum_{m,n\in\mathbb{Z}}}\Big(D^o_{(q_y,\frac{2m\pi}{d_c}),\frac{2n\pi}{d_c}}(b_{q_y,\frac{2m\pi}{d_c}}b_{-q_y,\frac{2n\pi}{d_c}}+b_{q_y,\frac{2m\pi}{d_c}}b^{\dagger}_{q_y,\frac{2n\pi}{d_c}}+b^{\dagger}_{q_y,\frac{2m\pi}{d_c}}b_{q_y,\frac{2n\pi}{d_c}}+b^{\dagger}_{q_y,\frac{2m\pi}{d_c}}b^{\dagger}_{-q_y,\frac{2n\pi}{d_c}})\\
&+&D^e_{(q_y,\frac{(2m+1)\pi}{d_c}),\frac{(2n+1)\pi}{d_c}}(b_{q_y,\frac{(2m+1)\pi}{d_c}}b_{-q_y,\frac{(2n+1)\pi}{d_c}}+b_{q_y,\frac{(2m+1)\pi}{d_c}}b^{\dagger}_{q_y,\frac{(2n+1)\pi}{d_c}}+b^{\dagger}_{q_y,\frac{(2m+1)\pi}{d_c}}b_{q_y,\frac{(2n+1)\pi}{d_c}}+b^{\dagger}_{q_y,\frac{(2m+1)\pi}{d_c}}b^{\dagger}_{-q_y,\frac{(2n+1)\pi}{d_c}})\Big).
\end{array}
\end{equation}
\end{widetext}
In the integration we have assumed that $\ell_y\gg \pi c/\omega_{TO}$ so that the Hamiltonian is approximately translationally invariant along the $\hat{y}$ direction.  The momentum dependent coefficients in $H_I^{(1)}$ corresponding to coupling with even and odd parity photonic fields are
\begin{equation}
\begin{array}{lll}
C^e_{(q_y,q_z),q'_z}&=&\hbar\omega_p\sqrt{\frac{ \omega_{TO}}{cd_cd_d\sqrt{\epsilon}}}\frac{1}{(q^2_z-q'^2_z)(q^2_y+q^2_z)^{1/4}}\times\\
& &(q_z\cos(d_d q'_z/2)\sin(d_d q_z/2)\\
& &-q'_z\cos(d_d q_z/2)\sin(d_d q'_z/2)),\\
C^o_{(q_y,q_z),q'_z}&=&\hbar\omega_p\sqrt{\frac{ \omega_{TO}}{cd_cd_d\sqrt{\epsilon}}}\frac{1}{(q^2_z-q'^2_z)(q^2_y+q^2_z)^{1/4}}\times\\
&&(q'_z\cos(d_d q'_z/2)\sin(d_d q_z/2)\\
& &-(q_z\cos(d_d q_z/2)\sin(d_d q'_z/2)).\\
\end{array}
\end{equation}
Similarly, for $H_I^{(2)}$ the momentum dependent coupling coefficients between even and odd field components is
\begin{equation}
\begin{array}{lll}
D^{e,o}_{(q_y,q_z),q'_z}&=&\frac{\hbar\omega_p^2}{4\sqrt{\epsilon}cd_c}\frac{1}{(q^2_y+q^2_z)^{1/4}(q^2_y+q'^2_z)^{1/4}}\\
&\times&\Big(\frac{\sin((q_z-q'_z)d_d/2)}{q_z-q'_z}\pm \frac{\sin((q_z+q'_z)d_d/2)}{q_z+q'_z}\Big)
\end{array}
\end{equation}
Note that the interaction is invariant under the parity transformation $\vec{q},\vec{q'}\rightarrow -\vec{q},-\vec{q'}$.  
The coupling coefficients are related by
\begin{equation}
\sum_{q_z'}[C^{e,o}_{(q_y,q_z),q'_z}]^2=\hbar \omega_{TO} D^{e,o}_{(q_y,q_z),q_z}.
\end{equation}  
In the bulk limit, $d_d\rightarrow d_c$, and for an infinite dielectric the momentum components satisfy $|q_z|=|q'_z|$.  Then we recover the appropriate infinite dielectric coupling relations (see e.g. \cite{16}):  
\[
\begin{array}{lll}
C^{e,o}_{(q_y,q_z),q'_z}&\rightarrow& \hbar\omega_{p}\sqrt{\frac{\omega_{TO}}{16c\sqrt{\epsilon_0}\sqrt{q^2_y+q^2_z}}},\\
D^{e,o}_{(q_y,q_z),q'_z}&\rightarrow&\frac{ \hbar\omega_{p}^2}{8c\sqrt{\epsilon_0}\sqrt{q^2_y+q^2_z}}.
\end{array}
\]
The total Hamiltonian for our system is then
\begin{equation}
\begin{array}{lll}
H&=&\displaystyle{\sum_{q_y}\Big[\sum_{n\in\mathbb{Z}}}\hbar \frac{c}{\sqrt{\epsilon}} \sqrt{q_y^2+(n\pi/d_c)^2}b^{\dagger}_{q_y,n\pi/d_c}b_{q_y,n\pi/d_c}\\
& &+\displaystyle{\sum_{q_z}}\hbar \omega_{TO}a^{\dagger}_{q_y,q_z}a_{q_y,q_z}\Big]+H_I^{(1)}+H_I^{(2)}.
\label{Htot}
\end{array}
\end{equation}

Since the Hamiltonian is a quadratic polynomial in field operators, $H$ can be diagonalized by a Tyablikov-Bogoluibov transformation \cite{26,23}.  The Hamiltonian is translationally invariant along $\hat{y}$ so different $q_y$ momentum components do not mix.   To obtain the solution, we introduce a new set of bosonic operators
\begin{equation}
\begin{array}{lll}
B_{q_y}&=&\displaystyle\sum_{\ell\in\mathbb{Z}}[X(q_y,\ell)b_{q_y,\ell \pi/d_c}+Z(q_y,\ell )b^{\dagger}_{-q_y,-\ell\pi/d_c}]\\
& &+\displaystyle{\sum_{q'_z}}[W(q_y,q'_z)a_{q_y,q'_z}+Y(q_y,q'_z)a^{\dagger}_{-q_y,-q'_z}].
\end{array}
\label{optrans}
\end{equation}
The coefficients of the expansion are determined by satisfying the commutation relation
\begin{equation}
[B_{q_y},H]=EB_{q_y}.
\end{equation}
The condition that a solution exists then reduces to the following dispersion relation which is an implicit function of energy: 
\begin{equation}
E^2-(\hbar\omega_{TO})^2=\displaystyle{\sum_{\ell\ \in\mathbb{Z}}}(f^e_{2\ell +1}(E)+f^o_{2\ell}(E)),
\end{equation}
where
\begin{equation}
\begin{array}{lll}
f^{e(o)}_{\ell}(E)&=&\frac{4\hbar \frac{c}{\sqrt{\epsilon}}\sqrt{q_y^2+(\ell \pi/d_c)^2}}{E^2-(\hbar c)^2(q_y^2+(\ell \pi/d_c)^2)/\epsilon}\Big((E^2-(\hbar\omega_{TO})^2)\times \\
&&D^{e(o)}_{(q_y,\frac{\pi\ell }{d_c}),\frac{\pi \ell}{d_c}}+\hbar\omega_{TO}\displaystyle{\sum_{q_z'}}[C^{e(o)}_{(q_y,\frac{\pi \ell}{d_c}),q_z'}]^2\Big)\\
&=&\frac{E^2(\hbar\omega_{p})^2d_d/2d_c}{\epsilon(E^2-(\hbar c)^2(q_y^2+(\ell \pi/d_c)^2)/\epsilon)}\big(1\pm\frac{\sin(\ell\pi d_d/d_c)}{\ell \pi d_d/d_c}\big).
\end{array}
\end{equation}
In the bulk limit, for each momentum wavevector $\vec{q}$ we obtain the appropriate dispersion relation \cite{16},
\begin{equation}
\Big(E^2-(\hbar\omega_{TO})^2\Big)\Big(E^2-\big(\frac{\hbar c|\vec{q}|}{\sqrt{\epsilon_0}}\big)^2\Big)-E^2\frac{(\hbar\omega_p)^2}{\epsilon_0}=0.
\end{equation}

When the cavity is tuned near the resonance condition at minimal separation, $d_c\approx \pi c/\omega_{TO}$, then the even parity $q_z=\pm\pi/d_c$ modes of the field are strongly coupled to the polarization current.  Restricting attention to the $q'_z=0$ modes of the polarization current which dominate the coupling and defining the creation operator for the even parity photon mode:
\begin{equation}
b^{\dagger}\equiv\frac{b^{\dagger}_{0,-\pi/d_c}+b^{\dagger}_{0,\pi/d_c}}{\sqrt{2}},
\end{equation}
and $a^{\dagger}\equiv a^{\dagger}_{0,0}$,
the interacting piece of the Hamiltonian then reduces to Eq. \ref{effHam}.


\begin{thebibliography}{10}

\bibitem{1}
Weisbuch, C., M. Nishioka, A. Ishikawa, and Y. Arakawa, 
\newblock Phys. Rev. Lett. 
\newblock{\bf 69}, 
\newblock 3314 (1992).

\bibitem{2}
J.P. Reithmaier, G. Sek, A. L\"{o}ffler, C. Hoffman, S. Kuhn, S. Reitzenstein, L.V. Keldysh, V.D. Kulakovskii, T.L. Reinecke, and A. Forchel, 
\newblock Nature {\bf 432}, 
\newblock 197 (2004); 
\newblock E. Peter, P. Senellart, D. Martrou, A. Lema\^itre, J. Hours, J.M. G\'erard, and J. Bloch, 
\newblock Phys. Rev. Lett {\bf 95}, 
\newblock 067401 (2004).

\bibitem{3}
G. Khitrova, H. M. Gibbs, F. Jahnke, M. Kira, and S. W. Koch, 
\newblock Rev. of Mod. Phys. {\bf 71}, 
\newblock 1591 (1999).

\bibitem{4}
D. Snoke, 
\newblock Science {\bf 298}, 
\newblock 1368 (2002).

\bibitem{5}
W. K\"utt, W. Albrecht, and H. Kurz,
\newblock IEEE J. Quantum Electron. {\bf 28},
\newblock 2434 (1992).

\bibitem{6}
M. C. Beard, G. M. Turner, and C. A. Schmuttenmaer, 
\newblock J. Phys. Chem. B {\bf 106}, 
\newblock 7146 (2002); 
\newblock P. Siegel, 
\newblock IEEE Trans. Microwave Theory Tech. {\bf 50}, 
\newblock 910 (2002).

\bibitem{7}
E.D. Murray, D.M. Fritz, J.K. Wahlstrand, S. Fahy and D.A. Reis, 
\newblock Phys Rev B {\bf 72},  
\newblock 060301 (2005).

\bibitem{8}
G.C. Cho, W. K\"utt, H. Kurz,
\newblock Phys. Rev. Lett. {\bf 65},
\newblock 764 (1990).

\bibitem{9}
M. Hase, M. Kitajima, S. Nakashima, and K. Mizoguchi, 
\newblock Phys. Rev. Lett. {\bf 88},  
\newblock 067401 (2002).

\bibitem{10}
G.A. Garret, A.G. Rojo, A.K. Sood, J.F. Whitaker and R. Merlin, 
\newblock Science {\bf 275},  
\newblock 1638 (1997).

\bibitem{11}
T. Feurer, J.C. Vaughan, K. A. Nelson, 
\newblock Science {\bf 299}, 
\newblock 374 (2003).

\bibitem{12}
O.V. Misochko, M. Hase, K. Ishioka, and M. Kitajima, 
\newblock Phys. Rev. Lett. {\bf 92}, 
\newblock 197401 (2004).

\bibitem{13}
A.V. Kuznetsov and C.J. Stanton, 
\newblock Phys. Rev. B {\bf 51}, 
\newblock 7555 (1995).

\bibitem{14}
K. Mizoguchi, O. Kojima, T. Furuichi, M. Nakayama, K. Akahane, N. Yamamoto, and N. Ohtani, 
\newblock Phys Rev. B {\bf 69}, 
\newblock 233302 (2004).

\bibitem{15}
M. Born and K. Huang, 
\newblock {\it Dynamical Theory of Crystal Lattices}
\newblock (Oxford University Press, Cambridge, 1954).

\bibitem{16}
J.J. Hopfield, 
\newblock Phys. Rev. {\bf 112}, 
\newblock 1555 (1958); 
\newblock D.L. Mills and E. Burstein, 
\newblock Rep. Prog. Phys. {\bf 37}, 
\newblock 817 (1974).

\bibitem{17}
T. Dekorsy, H. Auer, C. Waschke, H. J. Bakker, H. G. Roskos, H. Kurz, V. Wagner and P. Grosse, 
\newblock Phys. Rev. Lett. {\bf 74}, 
\newblock 738 (1995).

\bibitem{18}
Y.C. Shen, P.C. Upadhya, E.H. Linfield, H.E. Beere, and A.G. Davies, 
\newblock Phys. Rev. B {\bf69}, 
\newblock 235325 (2004).

\bibitem{19}
K. Mizoguchi, T. Furuichi, O. Kojima, M. Nakayama, S. Saito, A. Syouji, and K. Sakai, 
\newblock App. Phys. Lett. {\bf 87}, 
\newblock 093102 (2005).

\bibitem{20}
J. Hebling, G. Almasi, I.Z. Kozma and J. Kuhl, 
\newblock Opt. Expr. {\bf 10}, 
\newblock (1161) (2002).

\bibitem{21}
K.C. Huang, P. Bienstman, J.D. Joannopoulos, K.A. Nelson, and S. Fan, 
\newblock Phys. Rev. B {\bf 68}, 
\newblock 075209 (2003).

\bibitem{22}
D.W. Ward, J.D. Beers, T. Feurer, E.R. Statz, N.S. Stoyanov, and K.A. Nelson, 
\newblock Opt. Lett. {\bf 29}, 
\newblock 2671 (2004).

\bibitem{23}
V. Savona, Z. Hradil, A. Quattropani, and P. Schwendimann, 
\newblock Phys. Rev. B {\bf 49}, 
\newblock 8774 (1994).

\bibitem{24}
H.M. Lawler and E.L. Shirley, 
\newblock Phys. Rev. B {\bf 70}, 
\newblock 245209 (2004).

\bibitem{25}
L. C. Andreani and G. Panzarini, 
\newblock Phys. Rev. B {\bf 60}, 
\newblock 13276 (1999).

\bibitem{26}
V.M. Agranovich, Soviet Phys. JETP {\bf 37}, 307 (1960).


\end{thebibliography}
\end{document}